\documentclass[a4paper]{jpconf}
\usepackage{graphicx}

\def\PRL{Phys. Rev. Lett. }
\def\PRC{Phys. Rev. C }
\def\PRD{Phys. Rev. D }
\def\PLB{Phys. Lett. B }
\def\NPA{Nucl. Phys. A }

\newcommand{ \be }{\begin{equation}}
\newcommand{ \ee }{\end{equation}}
\newcommand{ \bea }{\begin{eqnarray}}
\newcommand{ \eea }{\end{eqnarray}}

\newcommand{ \ptt  }{p$_{\rm t}$~}
\usepackage{color}

\begin{document}
\title{Experimental Summary}

\author{Johanna Stachel        }
\address{Physikalisches Institut der
        Universit\"at Heidelberg, \\ 
        Philosophenweg 12,\\
        D69120 Heidelberg}

\section{Introduction}

This paper gives highlights of the experimental results shown at this
conference. As additional reference the four experimetal papers summarizing
the results from the first 3 years of RHIC running will be used
\cite{phenix,brahms,phobos,star}. This summary is a highly personal selection
of topical results and does not claim to be of comprehensive nature. The
author apologizes to all conference participants who contributed interesting
results that are not mentioned here.

\section{Global observables and initial condition}

Transverse energy production was studied by PHENIX and STAR
\cite{phenixet,staret} and measured as 600 GeV per unit pseudorapidity for the
most central 5 \% of Au + Au collisions (unless mentioned otherwise, here
always results at the top RHIC energy of $\sqrt{\rm s}$ = 200 GeV are
used). Using the Bjorken estimate and, very conservatively, a value $\tau_0$ =
1 fm/c gives for the initial energy density $\epsilon_0$ = 5.5
GeV/fm$^3$. Choosing instead the saturation scale as the measure for $\tau_0$
= 1/Q$_0$ = 0.14 fm/c gives a value 7 times larger. Most likely these two
values bracket the time scale of equilibration. In any case, the initial energy
density significantly exceeds the critical energy density for the phase
transition obtained by lattice QCD \cite{lattice_ec} as $\epsilon_{\rm c}$ =
0.7 GeV/fm$^3$.

Charged particle production is addressed by all four experiments. The total
charged particle multiplicity for the most central 5 \% of Au + Au collisions
at full RHIC energy was determined by BRAHMS~\cite{brahms_mult} to 4600. The
corresponding number measured by PHOBOS~\cite{phobos_mult} is 5300 for the
most central 6 \%. This leads to a total hadron multiplicity of about 7500. As
discussed by Roland~\cite{roland} at this conference, the concept of limiting
fragmentation holds very accurately for charged particle production over 3-4
units away from beam rapidity comparing SPS and RHIC
energies. Mohanty~\cite{mohanty} showed that photons (dominantly from $\pi^0$
decay) also exhibit this behavior. Assuming that this extends to LHC energy,
Roland~\cite{roland} estimated an upper limit of 1300 - 1800 for dN$_{\rm
ch}$/d$\eta$ for Pb + Pb collisions at LHC.

The centrality dependence of charged particle multiplicty distributions and,
in particular, its logarithmic growth with number of participants have been
taken as signs of gluon saturation and in that respect support the relevance
of the concept of color glass condensate at RHIC energies. Recent calculations
by Salgado~\cite{salgado} show that this approach works just as well for data
at top SPS energy and this casts some doubt whether color glass condensate is
indeed the correct underlying physical concept. In the talk of Hofmann the
scaling of charged particle multiplicity with beam energy was compared to
heavy ion data at lower beam energy as well as to pp and e$^+$e$^-$
data~\cite{hofmann,phobos}. The heavy ion data from top SPS energy upwards are
found to agree after participant scaling with pp data at twice the center of
mass energy. This is indicative of the large degree of stopping in a heavy ion
collisions. The energy deposition and therefore the energy available for
particle production is comparable to an e$^+$e$^-$ collision at the same
c.m. energy. This is reflected quantitatively in the BRAHMS data on projectile
nucleon rapidity shift as shown by Debbe~\cite{debbe,brahmseloss}. The best
estimate for the mean beam nucleon rapidity shift is 2.0 $\pm$ 0.4 units, very
close to the asymptotic value determined by Busza and
Goldhaber~\cite{buszagoldhaber} and later Busza and Ledoux~\cite{buszaledoux}
from central pA collisions. Of the incident beam energy 73 $\pm$ 6 \% is
available for heating of the fireball and particle production.

\section{Hadron production}

In the talk of Braun-Munzinger~\cite{pbm} it was shown, that RHIC data on
hadron production at mid-rapidity can be very accurately reproduced as the
particle yields of a chemically equilbrated system in terms of a grand
canonical ensemble (see \cite{hwareview} for a complete set of references). A
temperature of 177 $\pm$ 5 MeV is needed and this value is practically
unchanged from $\sqrt{\rm s}$ = 130 and 17.3 GeV, while the baryon chemical
potential drops continuously with increasing beam energy and has fallen, for
top RHIC energy, to 29 $\pm$ 5 MeV. It appears that from top SPS energy on a
constant temperature characterizes hadron yields, while at lower SPS energies
the temperature is significantly lower; at $\sqrt{\rm s}$ = 8.8 GeV it is only
148 $\pm$ 5 MeV.  A possible explanation how the system can achieve
equilibrium apparently in a very short time even for hadrons with multiple
(triple) strangeness and antinuclei was presented by
Braun-Munzinger~\cite{rapidequ}; due to the rapid increase in density in the
immediate vicinity of the phase transition multiparticle collisions become
very important and can drive even $\Omega$ yields into equilibrium within a
fraction of a fm/c.

The systematic behavior of hadro-chemical freeze-out parameters with beam
energy as shown by Braun-Munzinger~\cite{hwareview} leads to maxima of certain
particle ratios for intermediate beam energies driven by the interplay of
increasing temperature and decreasing baryon chemical potential; examples are
the ratio K$^+/\pi$ with a broad maximum around $\sqrt{\rm s_{nn}} \approx$ 10
GeV and the ratio $\Lambda/\pi$ with a pronounced peak at $\sqrt{\rm s_{nn}}$
= 5 GeV. In the talk of Friese (Fig.4 in~\cite{friese}) the measured
excitation function of such ratios by NA49 was shown and put into context with
lower and higher beam energy data. The observed ratio $\Lambda/\pi$ follows
the expected trend with the pronounced maximum with good precision. In the
ratio K$^+/\pi$ the data appear to show a sharper maximum and, in particular,
the NA49 data points at 80 and 158 A GeV/c beam momentum appear to lie nearly
two standard deviations below the thermal model prediction. While there has
been much speculation about the physics behind this deviation at this stage it
remains an experimental question requiring confirmation. It should be noted
that the ratios K$^-/\pi$ and K$^0_s$/h$^-$ do not show anything unusual.
 
\section{Azimuthal Anisotropies}

In has been observed for already quite some time that the elliptic flow
parameters v$_2$ as function of transverse momentum and for different particle
species can be remarkably well reproduced by hydrodynamics
calculations~\cite{huovinen,teaney,heinz}. The data are described
quantitatively up to about 2 GeV/c in transverse momentum. The elliptiv flow
observable for more massive particles like protons or Lambdas even exhibits
sensitivity to the equation of state favoring an equation of state with a QGP
- hadronic matter phase transition~\cite{huovinen}. As pointed out e.g. by
McLerran at this conference, it requires strong interactions at short times
for the hydrodynamic description to work so well. This could happen for
instance if the initial state is a color-glass condensate. Considering the
beam energy dependence of elliptic flow, the data show a continuous increase.
When plotting the elliptic flow coefficient divided by the eccentrity as
function of the charged particle rapidity density per cross sectional area,
the data appear to scale and to increase about linearly (see Fig. 25
in~\cite{flowna49}. The hydrodynamics results, conversely, show a weak drop
with beam energy and at top RHIC energy data and calculations meet. The
significant overprediction of data by hydrodynamics at SPS energy is linked to
the equation of state used, which has a softest point such that there is a
maximum in v$_2$ at SPS energy and a minimum at full RHIC energy~\cite{star}.
As shown by Jacak (Fig.4 in~\cite{jacak}) the elliptic flow increases in data
by about 50\% from SPS energy to $\sqrt{\rm s_{nn}}$ = 62.4 GeV and then
remains about constant for larger beam energies althoug the energy density
there increases about 30 \%, supporting a soft equation of state. In the talk
by Jacak were also shown the results of various hydrodynamics calculations
together with transverse momentum spectra and elliptic flow for protons and
pions (Fig.5 in~\cite{jacak}). While qualitatively all calculations that
incorporate a phase transition between QGP and hadronic phase reproduce the
overall features shown by the data, there are significant deviations in
detail. This has to do with how the hadronic phase and freeze-out are treated.
Common to all calculations is a rapid equilibration on a time scale of 0.5-1
fm/c.

Another type of scaling has been discovered in the elliptic flow data at RHIC:
when normalizing both the flow coefficient v$_2$ and the transverse momentum
to the number of constituent quarks in a hadron, the data for different
hadronic species scale for transverse momenta above about 2 GeV/c as shown by
Xu (Fig.3 in~\cite{xu}). This has been interpreted as partons in the QGP
carrying the flow information and preserving it in coalescing~\cite{xu}.

\section{Jet quenching}

One of the highlights of the RHIC program is the observation of the
suppression of production of hadrons at high transverse momentum. This was
addressed in several presentations at this conference. When normalizing the
transverse momentum spectra observed in Au + Au collisions to the
corresponding pp spectra multiplied with the number of binary collisions, a
suppression factor R${\rm_{AA}(p_t)}$ is obtained. Similarly central and
peripheral Au + Au collisions can be compared. It is observed in all four
experiments~\cite{phenix,brahms,phobos,star} that for central Au + Au
collisions production of pions/charged hadrons is suppressed by a factor 4-5
for transverse momenta in excess of 4 GeV/c. There is reasonable agreement
between the experiments, although around 2.5 GeV/c a 20 \% difference between
the STAR data and those of the other experiments is noted (see
e.g. Fig.2 of \cite{hemmick}). Since the suppression is not seen in d + Au
collisions~\cite{phenix,brahms,phobos,star}, it is obviously an effect induced
by the medium the parton or jet traverses.

Presumably the suppression is due to medium induced radiative energy loss of
the parton before it fragments and hadronizes as was
predicted~\cite{wang,baier}. Calculations employing a large initial gluon
rapidity density of about 1100 can account for the data~\cite{vitev}.  The
beam energy dependence of the R${\rm_{AA}}$ factor was presented recently by
d'Enterria~\cite{denterria} and it appears that the suppression evolves in a
very smooth way from top SPS energy onwards. The R${\rm_{AA}(p_t=4 GeV/c)}$
value of 1.0 measured at the SPS represents already a slight suppression as
compared to the normal Cronin enhancement~\cite{denterria}.  Going from
$\sqrt{\rm s_{nn}}$ = 17.3 to 62.4 to 200 GeV the gluon rapidity density
needed grows from 400 to 650 to 1100~\cite{vitev}. Alternatively this
suppression can be described by increasing and large opacities of the medium
traversed~\cite{wangwang}.

These high initial gluon densities correspond to an initial temperature of
about twice the critical temperature and to initial energy densities
$\epsilon_0$ = 14 - 20 GeV/fm$^3$ well in line with the initial conditions
needed for the hydrodynamics calculations to describe spectra and elliptic
flow (see section 4) and bracketed by the estimates based on the
Bjorken formula (see section 2).

It was observed since some time that the high \ptt suppression pattern is
different for different hadronic species. This is very cleanly demonstrated by
recent STAR data on the ratio $\Lambda$/K$^0_s$ which was shown by
Lamont~\cite{lamont} as function of \ptt for different centralities. For
central collisions this ratio peaks at \ptt $\approx$ 3 GeV/c at a value of
1.6, close to the ratio 3/2 expected in quark coalescence models. However,
such models at present give quantitatively only a rough description of the
data as shown in Fig. 2 of ~\cite{lamont}. In the talk of Jacak a similar rise
of the ratio for proton over pion \ptt spectra is shown for d + Au collisions
(see Fig. 2 in~\cite{jacak}) and the question arises why (constituent) quark
coalescence would be an appropriate description for a d + Au collision.  To
test the coalescence hypothesis jet partners have been looked for for a high
\ptt meson or baryon both on the near and on the away side. The fact that
within present measurement accuracy a jet partner is equally likely for a
trigger baryon or meson~\cite{phenixjet} (very similar data from STAR where
shown by Mischke and Mironov at this conference) does not strengthen the
valence quark coalescence scenario.

In d + Au collisions no suppression is seen in central collisions at
mid-rapidity. Recent data by BRAHMS on R${\rm_{dAu}}$ as function of
pseudo-rapidity shown that a significant suppression sets in at forward
rapidities~\cite{brahmsforward}. This is supported by similar data from
PHOBOS~\cite{phobosforward}. It has been suggested that this could be a
manifestation of the color glass condensate visible as one probes smaller
values of x and a corresponding analysis describes the data
well~\cite{cgc}. On the other hand, the same data can be described equally
well in terms of the parton coalescence picture~\cite{hwa}. In the talk by
Vitev it was shown, that the same feature is exhibited by d + Au data at
$\sqrt {\rm s}$ = 19.4 GeV. This sheds doubt on both the color glass and the
quark coalescence descriptions and points to alternative explanations such as
put forward by Kopeliovich et al.~\cite{kopel} e.g. in terms of energy loss of
the incoming parton.

Parton thermalization is displayed in a very clean way be recent results of
the STAR collaboration~\cite{startherm,mischke}. Evaluating the mean
transverse momentum in a cone opposite to a high \ptt trigger particle as a
funtion of centrality, a gradual decrease for more central Au + Au collisions
is observed and in the most central collisions a value very close to the
inclusive mean \ptt is reached (see Fig. 2 in~\cite{mischke}). To actually
connect the high energy parton energy loss to the radiated gluons Vitev
calculated di-hadron correlations (see Fig. 3 in~\cite{vitev}). Comparing
calculations with and without the socalled gluon feed-back to STAR
data~\cite{startherm} shows that quenching of the away-side jet at high \ptt
is accompanied by a simultaneous strong enhancement at \ptt below 2 GeV/c.

In azimuthal correlations of two high \ptt particles it was seen that the
away-side peak disappears in central Au + Au collisions. Recently it was shown
that the effect is very strong in case the away-side jet is emitted out of the
reaction plane and much weaker for emission in the reaction
plane~\cite{starjet}. This supports the strong correlation of the suppression
with the length of matter traversed by the parton. When lowering the \ptt cut
on the correlated hadron, a very broad structure appears on the side opposite
to the trigger particle. This was shown by Mischke for a cut on the correlated
hadron of \ptt = 0.15 - 4. GeV/c (see Fig. 1 in~\cite{mischke}). This calls to
mind a similar observation at SPS energy by CERES~\cite{ceres} where for a
condition \ptt $\geq$ 1.2 GeV/c also very strong broadening of the away-side
structure with increasing collision centrality was observed. In the talk of
Jacak a tantalizing observation was shown~\cite{phenixhole}: For a trigger
particle \ptt of 4-6 GeV/c and a correlated particle \ptt of 1.0 - 2.5 GeV/c
the away-side peak seen in peripheral Au + Au collisions develops actually
into a hole at $\Delta \phi = \pi$ for more central collisions while a very
broad peak appears with a maximum at $\Delta \phi = \pi-1$ as can be seen in
Fig. 8 of~\cite{jacak}. Could this be the Mach cone due to the sonic boom of
the quenched jet? This goes back to an idea developed by St\"ocker et al. for
nucleons and nuclei emitted in nuclear reactions in the mid 1970ies (see
references [14,15] in~\cite{stoeckersum}) and was suggested for the current
scenario of a parton traversing a quark-gluon
plasma~\cite{stoecker,casalderrey}. If this could be established it would have
far reaching consequences since it would be an observable linked directly to
the speed of sounds of the quark-gluon plasma and thereby its equation of
state. At present it remains an experimental challenge to establish an actual
cone topology in two dimensions.

\section{Open charm and charmonia}

Open charm has been measured indirectly from the inclusive electron \ptt
spectra after subtracting known contributions from photon conversions and
light hadron decays by PHENIX~\cite{phenixcharm}. Recently similar
measurements have become possible at high \ptt for STAR using the calorimeter~
\cite{lamont}.  The spectrum remaining after subtraction is
dominated~\footnote{A possible contribution to the electron spectrum from the
Drell-Yan production process cannot be ruled out at present, though.} by open
charm and beauty contributions but the method is limited by the systematic
error introduced by the subtraction. A significant step ahead will come in the
future when silicon micro-vertex-detectors will be implemented.  Nevertheless,
very interesting results have come already from the present method. In the
talk of Jacak results for an elliptic flow analysis of the open charm decay
elecrons were shown (see Fig. 6 of~\cite{jacak}). There is a significant
nonzero value in the transverse momentum range 0.4 - 1.6 GeV/c. At this
meeting STAR data were shown by Lamont~\cite{lamont} that complement the
PHENIX results and extend the overall transverse momentum coverage by adding
the range \ptt = 1.5 - 3.0 GeV/c. Together the data paint a consistent picture
that indeed the electrons from open charm decay exhibit elliptic flow. What is
more, the data are described quantitatively by a calculation~\cite{greco}
based on thermalization and elliptic flow of the charm quark (see Fig. 7
of~\cite{lamont}); the corresponding calculation with no charm quark flow (and
hence only an effect from the flow of the light quark in a D meson) is a
factor two below the data, altough with the presently still large error bars
this is an effect of only a few standard deviations. The present data can be
taken as an indication that the charm quark thermalizes to a significant
degree. Note that this is a necessary prerequisite for any formation of
charmed hadrons by statistical hadronization~\cite{statcharm}. On the other
hand, in that case also jet quenching should be observed for charmed
hadrons. Indeed in the talks by Hamagaki and Tabaru~\cite{tabaru} it was shown
that electron spectra after the subtraction of contributions from conversion
and light hadron decays show high \ptt suppression for central collisions. The
R${\rm _{AA}}$ factor drops practically as low as for pions at \ptt of 4
GeV/c, i.e. to values of about 0.2 (see Fig.7 in~\cite{jacak}). In a recent
publication~\cite{armesto} the suppression for electrons from D meson decay
was studied for different transport coefficients and in the talk of Jacak it
was shown, that the present data would be consistent with a calculation using
a transport coefficient of 14 GeV$^2$/fm (see Fig. 2 of~\cite{armesto}) at the
upper end of the range needed to reproduce the data for pions. This is very
surprizing, in particular also in view of the fact, that at \ptt of about 4
GeV/c also the contribution of b-quarks to the electron spectrum should become
sizeable.
 
The overall charm production cross section at RHIC energy has been measured
indirectly by PHENIX~\cite{phenixcharm} for $\sqrt{\rm s}$ = 130 and 200 GeV
from the inclusive electron spectra in the way described above for Au + Au
collisions and at full RHIC energy also for d + Au and pp
collisions~\cite{phenixcharm2}. It is found that the integrated charm cross
section, when scaled with the number of binary collisions, agrees for all
three collisions systems~\cite{tabaru}. The value is about 30\% above a NLO
pQCD calculation (as shown e.g. by Lamont~\cite{lamont} in these proceedings)
but agrees within errors. In STAR, D mesons have been reconstructed via their
hadronic decay to K$\pi$ in d + Au collisions and a charm cross section per
nucleon nucleon collision has been extracted~\cite{starcharm}. It is twice as
large as the PHENIX value by about two standard deviations. This obvious
experimental discrepancy needs to be resolved; most likely this will come only
with the future charm measurement using the displaced vertex feature.

For charmonia there were final results for NA50 at the CERN SPS presented at
this meeting by Quintans~\cite{quintans}. Analysing all pA data, a cross
section for normal nuclear absorption of 4.1$\pm$0.4 mb was
extraced~\cite{na50}. To this normal nuclear suppression all results from
heavy ion collisions can be compared. It turns out that S + U data as well as
data from peripheral Pb + Pb collisions agree with this normal nuclear
absorption curve. For transverse energies above 40 GeV or a length of nuclear
matter seen by the J/$\psi$ of L $\geq$ 7 fm the points from Pb + Pb collisions
fall increasingly below this normal nuclear absorption curve. The range of L =
5.3 - 8 fm covered by the new NA60 experiment for In + In collisions is
exactly in the interesting region where the deviation is starting. First data
were shown by Seixas at this meeting (although not contained in these
proceedings) and they confirm well the systematic trend seen in Pb + Pb
collisions. Nothing has changed on the front of theoretical
interpretation. The suppression can be explained by disappearance of the
J/$\psi$ (or possibly only the charmonia states that feed it) in a hot colored
medium or by interaction with comovers, albeit with a very large density of
more than 1/fm$^3$, i.e. a value not deemed achievable for a hadron gas.

\section{Direct photons}

Direct photons in pp collisions have been measured at top RHIC energy by
PHENIX~\cite{photonspp}. It was shown at this conference by Cole and Bathe
(see Fig. 1 in~\cite{bathe}) that they agree well with a NLO pQCD calculation
by Gordon and Volgelsang~\cite{vogelsang}. The data are still statistics
limited plus there is some remaining scale uncertainty. Here additional high
luminosity pp running is desirable. For Au + Au collisions, the measurement by
PHENIX~\cite{photonsauau} shown at this meeting by Bathe and Jacak agrees also
well with the same NLO pQCD calculation. There is no definitive answer yet
whether k$_{\rm t}$ broadening is needed. The centrality dependence is well
reproduced over the entire range in \ptt covered out to 13 GeV/c just by
scaling with the number of collisions (see Figs. 2,3
in~\cite{bathe}). Comparing this to the observed high \ptt suppression seen
for hadron spectra (see above) confirms that this is indeed a final state
effect due to the medium that should not and apparently does not affect
photons. Comparing the measured $\gamma / \pi^0$ ratio to the expected one due
to hadronic decays a direct photon component is apparent for \ptt above 4
GeV/c. With systematic errors of 10 \% this means that at the present stage
there is unfortunately no sensitivity yet to thermal photons radiated by the
plasma; they are expected to dominate over direct hard photons below about 3
GeV/c. To achieve the required sensitivity of a few \% in $\gamma / \pi^0$ in
the range \ptt = 1-3 GeV/c requires a quantum jump in the quality of the
experimental data. Nevertheless, this very difficult measurement is very
desirable.

\section{Fluctuations and correlations}

For this interesting subject I refer to the authoritative review given by
Voloshin in these proceedings~\cite{voloshin}. Here only very few points are
picked up. Transverse momentum fluctuations have been studied at the SPS and
at RHIC and many different observables have been proposed and have been used
to quantify these fluctuations. Nonstatistical fluctuations have been seen at
all energies from $\sqrt {\rm s_{nn}}$ = 8 to 200 GeV. Voloshin has proposed
to use the momentum correlator as a measure. In this quantity all present
measurements give a relative fluctuation of 1 \% without any hint to increased
fluctuations possibly indicating the vicinity of a critical point (see Fig. 4
of~\cite{voloshin}). The question was raised by Koch at this meeting whether
possibly the data cover not the optimal region in \ptt. Another possibility is
that the data integrated over opening angles average over too many different
contributions to correlations which might lead to a loss in
sensitivity~\cite{misko}. Voloshin pointed out that, nevertheless, the current
data are very useful because they are sensitive to velocity profiles for the
expansion of the fireball as demonstrated in his Fig.5. This should indeed be
exploited more to get quantitative results that then constrain the
interpretation of transverse momentum spectra and two particle correlations.

A very tantalizing preliminary NA49 result on the charged kaon to pion
multiplicity fluctuations was shown by Friese (see Fig.6 in~\cite{friese}): at
the smallest SPS energies the fluctuations are larger than expected in
contrast to the fluctuations in the proton to pion ratio that look
'normal'. It would be extremely desirable to confirm this very difficult and,
if confirmed, possibly far reaching measurement with better particle
identification capability.

{\it Acknowledgements.}  Valuable discussions with P.~Braun-Munzinger are
gratefully acknowledged.  This work was supported in part by the HGF GSI
virtual institute VH-VI-146.

 \end{document}